\documentstyle[12pt,titlepage,fleqn]{article}

\newcommand{\be}{\begin{equation}}
\newcommand{\ee}{\end{equation}}
\newcommand{\ba}{\begin{eqnarray}}
\newcommand{\ea}{\end{eqnarray}}

\newcommand{\om}{\omega}

\newcommand{\non}{\nonumber}
\newcommand{\oh}{\frac{1}{2}}
\newcommand{\IIA}{type IIA superstring}

\begin{document}

\begin{titlepage}
\vspace*{0.1truecm}
\begin{center}

{\large\bf  HETEROTIC AND TYPE I STRINGS FROM TWISTED SUPERMEMBRANES}

\vspace{1cm}

{\large Fermin ALDABE\footnote{E-mail: faldabe@phys.ualberta.ca}}

\vspace{1cm}

{\large\em Theoretical Physics Institute,
University of Alberta\\
Edmonton, Alberta, Canada, T6G 2J1}

\vspace{.5in}

\today \\

\vspace{1cm}
{\bf ABSTRACT}\\
\begin{quotation}

\baselineskip=1.5em

As shown by Ho\v{r}ava and Witten, there are gravitational anomalies at
the boundaries of $M^{10}\times S^1/Z_2$ of 11 dimensional supergravity. 
They showed that only 10 dimensional vector multiplets belonging to $E_8$
gauge group can be consistently coupled to this theory. Thus, the
dimensional reduction of this theory should be the low energy limit of the
$E_8\times E_8$ heterotic string. Here we assume that M-theory is a theory
of supermembranes which includes twisted supermembranes. We show that for
a target space $M^{10}\times S^1/Z_2$, in the limit in which $S^1/Z_2$ is
small, the effective action is the $E_8\times E_8$ heterotic string.  We
also consider supermembranes on $M^{9}\times S^1\times S^1/Z_2$ and find
the dualities expected from 11 dimensional supergravity on this manifold.
We show that the requirements for worldsheet anomaly cancellations at the
boundaries of the worldvolume action are the same requirements imposed on
the Ho\v{r}ava-Witten action.

\end{quotation}
\end{center}

\end{titlepage}

\section{Introduction}

There is strong evidence for the conjecture
that all string theories can be derived from M-theory, which has
as its low energy limit 11 dimensional supergavity.  In 
\cite{D}, it was shown that the type IIA string can be obtained from
a double dimensional reduction of the supermembrane whose worldvolume is
a torus propagating in worldvolume time.  Townsend \cite{T} has shown that
the complete spectrum of the \IIA\ can be obtained from the supermembrane
after identification of the solitonic string with the fundamental string
and the solitonic 11 dimensional membrane with the fundamental membrane.
In \cite{T1,HT}, it was shown that 11 dimensional supergravity on K3
is dual to heterotic string on $T^3$.
In \cite{W1}, it has been shown that the strong coupling limit of \IIA\
yields 11 dimensional supergravity.  Also, the spectrum of
Type IIB string on an $S^1$ can be identified with that of 11 dimensional 
supergravity on $T^2$ \cite{Sc}.  This evidence is also supplemented 
by other duality relations between string theories in various dimensions
(see for example \cite{W1,Sen,ASY}).

As explained by Ho\v{r}ava and Witten \cite{HW},
the gravitational anomaly will be non vanishing   
for the 11 dimensional supergravity action on $M^{10}\times S^1/Z_2$.
The orbifold $S^1/Z_2$ has two fixed points where the manifold is 
singular because there is no tangent bundle defined there.
Then, there will be
contributions to the gravitational anomaly of the 11 dimensional 
supergravity action exactly at the fixed points.  This anomaly is
just the 10 dimensional gravitational anomaly.  In order to have an
anomaly free theory,
there must also be additional massless states which are vector 
multiplets in the twisted sectors of the M-theory which
can make the total anomaly, gravitational and gauge, vanish.  

The gravitational anomaly in ten dimensions can be canceled by the addition
of 496 vector multiplets.  This means that the gauge groups which can be
considered for the anomaly cancellation must have dimension 496.  The 
possible gauge groups which arise are SO(32) and $E_8\times E_8$.
However, The gauge group $E_8\times E_8$ is the 
only candidate, because there are
two hyperplanes, determined by the fixed point of the 
orbifold
\cite{HW}, each contributing symmetrically to the anomaly. Then,
248 massless vector multiplets must propagate on each 
hyperplane.
The presence of these  two hyperplanes forbids the SO(32) group because
the vector multiplets can only be placed in one or the other plane.

In fact, in \cite{HW1}, Ho\v{r}ava and Witten showed that only 
$E_8\times E_8$ 10 dimensional super Yang-Mills can be consistently
coupled to 11 dimensional supergravity.  Such a coupling
requires a modified Bianchi identify for the 4-form field strength of
11 dimensional supergravity.  In turn, this 
leads to non trivial gauge transformations
of the 3-form fields of 11 dimensional supergravity.

The natural question, is how the vector multiplets are generated.  To answer
this question we must have some knowledge about the M-theory.  As of yet we
do not know which is the theory that has as its low energy limit 11 
dimensional supergravity. However, there is speculation
\cite{TBS,T,HT,Sc} that the
low energy limit of the supermembrane action is 11 dimensional supergravity.
Here we make this assumption to test whether the supermembrane can
reproduce the results obtained for 11 dimensional supergravity on $S^1/Z_2$.
The motivation for such an assumption is the work of Duff et. al. \cite{D}
where it was shown that the double dimensional reduction of the
supermembrane is the type IIA string.

In the next  section,
we review the three  dimensional anomaly for the 
twisted solitonic membrane which should be identified with the twisted
 supermembrane and write the anomaly free action for the twisted supermembrane.
The anomaly free conditions are shown to be the same ones as those
required by the Ho\v{r}ava-Witten action \cite{HW1}.
Double dimensional reduction of this action on $S^1/Z_2$ is shown to yield 
the heterotic string with $E_8\times E_8$ gauge group.
In section 3 we consider the double dimensional reduction of the
twisted supermembrane action on $S^1\times S^1/Z_2$ to show that the
dualities of 11 dimensional supergravity
on $M^9\times 
S^1\times S^1/Z_2$  \cite{HW}
follow from the choice of dimensional reduction.
The last section is devoted to the discussions.  We propose that
membrane/string duality in seven dimensions \cite{T1}, which is responsible 
for string/string duality in six dimensions is a duality between 
twisted and close membranes in seven dimensions.  We also suggest that 
the twisted membranes, which have current algebras at its boundaries,
is the candidate for the construction of D-branes in M-theory.

\section{Heterotic String Form The Supermembrane }

As argued by Townsend \cite{T},
we must identify the solitonic membrane with the fundamental membrane 
of 11 dimensional supergravity in order
to derive the type IIA spectrum from the supermembrane spectrum. An additional
reason for such an identification is the believe that the supermembrane action
has a continuous spectrum 
\cite{DW}.  The spectrum is continuous because there is no energy needed to 
deform the membrane to 
create spikes of zero area.
A way to truncate the spectrum, as suggested in \cite{T}, 
 is to identify the supermembrane with
the solitonic membrane, thus, the fundamental membrane  inherits
 a thickness due to gravitational effects.  This was shown to be
the case in  \cite{AL}. 
Due to the presence of the
thickness term, spikes will necessarily have non vanishing area and will
require energy to be created.
This identification also 
allows us to obtain the properties of the supermembrane
simply by studying the properties of the solitonic membrane.

In \cite{HW}, the supersymmetry properties of 
a stable solitonic  membrane on ${M}^{10}\times S^1/Z_2$ where studied 
corresponding to $x^2=...=x^9=0$
with $x^1$ as time , and $x^{11}$ a coordinate on
the orbifold $S^1/Z_2\simeq I$.  It was shown that 
the  unbroken supersymmetries are given by spinors 
$\epsilon$ satisfying
\ba
\Gamma^{11}\epsilon&=&\epsilon\non\\
\Gamma^{1}\Gamma^{10} \epsilon&=&\epsilon.\label{bc2}
\ea
Then, at the boundaries which are given by the fixed points of the
orbifold, there will be left and right moving bosons and right moving fermions,
but no left moving fermions. We may then expect the same boundary conditions
in the worldvolume of the supermembrane.

As explained in \cite{HW}, the action of the solitonic membrane can have
gravitational anomalies.  A three dimensional action  will in general
not have these anomalies, but because there are boundaries, there will
be  effective two dimensional field theories defined at the boundaries.
This in turn can give rise to gravitational anomalies much in the same manner
there are 10 dimensional gravitational anomalies 
present in 11 dimensional supergravity
theories.  
In addition, there is also a possibility of having gauge anomalies.

Surprisingly, we find that the worldsheet gravitational and gauge anomaly 
cancellation requires a modified Bianchi  identity \cite{HT1}
which is of the same
form as the modified Bianchi identity of \cite{HW1} required to 
couple 11 dimensional supergravity to 10 dimensional super Yang-Mills.  
Consider  a world volume given by {\bf R}$\times S^1\times I$.
The motivation for such a choice of worldvolume is based on the work of 
Ho\v{r}ava who constructed the type IA string from worldsheet orbifolds
\cite{H1,H2}.
In addition, the presence of  $I$ will allow for the double
dimensional reduction leading to a heterotic string.  
The supermembrane action in the bulk (excluding the boundary of the worldvolume)
 of the
theory will be \cite{TBS}
\be
S_M=\int d^3\zeta (\sqrt{-det(\hat{E}^a_i\hat{E}^b_j
\eta_{ab})}+\frac{1}{6}
\epsilon^{ijk}\hat{E}^A_i\hat{E}^B_j\hat{E}^C_kB_{CBA}).\label{s}
\ee
$E^A_i=\partial_i \hat{Z}^N \hat{E}^A_N$ and $\hat{E}^A_M$ is a superelfbein
and $B_{CAB}$ is an antisymmetric 3-tensor.
 $Z^N=(x^n,\theta^{\nu}),\ n=1,...,10$ are supercoordinates on $M^{10}$, 
 and $Z^{11}$ 
is a supercoordinate
on $I$.  The coordinates $\zeta^i$ are worldvolume coordinates.
 The worldvolume part $S^1$ will have as coordinate $\sigma$.  The worldvolume 
part
$I$ will have as coordinate $\rho$.  
The world volume is then a cylinder propagating in world volume time, $\tau$,
 which is a coordinate on ${\bf R}$. The ends of the cylinder are at 
$\rho\pm a$. 

Using the reparametrization invariances of the action $S_M$ we set 
\be
\rho=x^{11}\label{rep}.
\ee
The boundary will be independent of $\rho$.  Thus,
the field $Z_{11}$ does not 
contribute to the gravitational anomaly
because at the boundary it has no dynamics.

We must also impose the condition
that the fermions which are left moving at the boundaries vanish
at the boundaries in order to satisfy (\ref{bc2}):
if
\be
\partial_+\theta^{A \alpha}(\tau,\sigma,\pm a)=0
\ee
where $\partial_{\pm}=\oh(\partial_{\tau}\pm \partial_{\sigma})$,
then
\be
\theta^{A \alpha}(\tau,\sigma,\pm a)=0.
\ee
The index $\alpha$ above denotes an SO(8)
spacetime spinor while the index $A=1$
labels right movers at the boundary
and the index $A=2$ labels left movers at the boundary \cite{GSW}.  
Therefore, condition (\ref{bc2}) implies that 
\be
\theta^{2\alpha}(\tau,\sigma,\pm a)=0.
\ee
With this boundary conditions, the action for the twisted 
membrane takes the form
\be
S_M
+ \int_{\partial M^3} 
(\sqrt{-det({E}^a_i{E}^b_j
\eta_{ab})}+\frac{1}{2}
\epsilon^{ij}E^A_i{E}^B_jB_{BA})
.\label{s5}
\ee
where $\partial M^3$ is two copies of $S^1\times {\bf R}$, and the 
superzenbeins at the boundaries are
\be
E^{m}_i=\partial_i X^{m}-i\bar{\theta}^{1\alpha}
\Gamma^{m}\partial_i\theta^{1\alpha}\label{ocho}
\ee

The action (\ref{s}) has a worldvolume $\kappa$ symmetry which insures
spacetime supersymmetry in the bulk of $M^{10}\times I$.
This $\kappa$ symmetry has been shown to 
be dimensionally reduced  to a worldsheet $\kappa$ symmetry of the type IIA
string
\cite{D}, so that the 
$\kappa$ transformations at the boundary are
\ba
\delta \theta^A&=&2i \Gamma\cdot(\partial_iX^{m}-i\bar{
\theta}^A\Gamma^{m}\partial_i
\theta^A)\kappa^{A i}\\
\delta X^{m}&=&i\bar{\theta}^A\Gamma^{m}\delta\theta^A.
\ea
However, the boundary condition (\ref{bc2}) projects this $\kappa$ 
transformations to those of action (\ref{s5}) which take the form
\ba
\delta \theta^1&=&2i \Gamma\cdot(\partial_iX^{m}-i\bar{
\theta}^1\Gamma^{m}\partial_i
\theta^1)\kappa^{1 i}\\
\delta X^{m}&=&i\bar{\theta}^1\Gamma^{m}\delta\theta^1.
\ea
Thus, the action (\ref{s5})
at the boundary has the $\kappa$ symmetry of the heterotic
string.  

Alternatively,  it is possible to write the
two dimensional action at the boundary in the NSR formalism
in terms of Fermi superfields using
(0,1) superspace which guaranties the existence of a $\kappa$ symmetry
required by N=1 spacetime supersymmetry \cite{HT1}.  
This is possible because
the NSR fermions are vectors of SO(8) while the fermions of the Green-Schwarz
formalism are SO(8) spinors.  
Using the triality properties of SO(8), it
is possible to map the Green-Schwarz spinors into NRS fermions transforming
as vectors of SO(8).  The action (\ref{s5}) then is the sum of two terms.
One term  is the Green-Schwarz supermembrane action,  in addition there
is an NRS heterotic string (without gauge group) 
\ba
S&=&S_M+\int_{\partial M^3}
\{\oh (G_{mn}\eta^{ij}+B_{mn}\epsilon^{ij})\partial_ix^m\partial_jx^n\non\\
&&+\oh i \lambda^a_+[\partial_- \lambda^a_+ 
+\om_m^{(+)ab}\partial_-x^m\lambda^b_+]\}\label{s6}
\ea
where $G$ and $B$ are the background metric and antisymmetric tensor; 
the latter is obtained after dimensional reduction of the 3-form tensor of
the supermembrane action. $\eta$
is the flat worldsheet metric and we use as in \cite{HT1} the orthonormal
frames $E^a_n$ such that the right moving Majorana-Weyl fermions satisfy
$\lambda^a=E^a_{n}\theta^{n}$, and we have used the triality properties of
SO(8) to map the SO(8) spacetime spinor $\theta^{\nu}$ to an SO(8)
vector $\theta^n$.

In addition, we may add an $E_8$  current algebra at each boundary.  
As shown in \cite{HW1}, the 10 dimensional
Yang-Mills supermultiplet at each boundary 
which couples to 11 dimensional supergravity
on a manifold $M^{10}\times I$ must be in the adjoint of the 
$E_8$ gauge group.  
This means that after using (\ref{rep}), the action (\ref{s5}) requires 
the addition of an $E_8$ current algebra at each boundary.   
Thus the total action we will consider is a sum of two terms: a bulk term and
a boundary term.  The bulk term will be the usual supermembrane action of
\cite{TBS}; the boundary term will be the non linear sigma model with
(0,1) supersymmetry and $E_8$ gauge group.  It will have a conformal anomaly,
however, the bulk theory is already not conformally invariant.  The explicit
form of the boundary term can be obtained from the action considered in
\cite{HT1} after dropping half of the left moving fermions
so that the remaining left moving fermions only couple to one $E_8$

\ba
S&=&S_M+\int_{\partial M^3}
\{\oh (G_{mn}\eta^{ij}+B_{mn}\epsilon^{ij})\partial_ix^m\partial_jx^n\non\\
&&+\oh i \lambda^a_+[\partial_- \lambda^a_+ 
+\om_m^{(+)ab}\partial_-x^m\lambda^b_+]
\non\\
&&+\oh i \psi^A_-[\partial_+ \psi^A_++A_{n}^{AB}\partial_+x^n\psi^B_-]
\non\\
&&+
\frac{1}{4}F_{abAB}\lambda^a_+\lambda^b_+\psi^A_-\psi^B_-\}
,\label{S}
\ea
The left moving Majorana-Weyl fermions
$\psi^A$ are also in the NRS formalism and
only live on the boundaries; they couple naturally to 
$A^{AB}$  and $F_{abAB}$,
the Yang-Mills connections and field strength respectively, where
the indices $A$ and
$B$ run over the representation of one $E_8$ only.
Recall that the boundary $\partial M^3$ is two copies of $R\times S^1$.

{\em Thus, the closed supermembrane couples naturally to the 
11 dimensional supergravity multiplet, and we expect that the twisted membrane
(\ref{S})
should similarly couple to the bulk supergravity with $E_8$ Yang-Mills
super multiplets propagating at the spacetime boundary.}

Action (\ref{S}) will have gravitational and gauge anomalies given by the
two dimensional theory.  Fortunately, these anomalies have been computed
for non-linear sigma models with (0,1) superspace \cite{HT1}.  Consider
the total contribution of the anomaly to be formally described by
\be
A(R)+B(R,F_1)+B(R,F_2)\label{uno}
\ee
where $A(R)$ is the contribution from general coordinate invariance (
comming from the right
moving fermions at the boundary) and 
$B(R,F)$ is the contribution  to the 
gauge anomaly (comming from the left moving fermions). Motivated by \cite{
HW,HW1}
we introduce  the quantity
\be
\oh A(R)+B(R,F)\label{dos}
\ee
at each boundary,
so that (\ref{uno}) is the sum of the two boundary contributions (\ref{dos}).
Therefore, in computing the gravitational and gauge anomalies,  each
boundary will contribute half of its usual 
gravitational anomaly and will contribute its full amount to the
gauge anomaly comming from an $E_8$ multiplet.

 The 
derivation of the anomalies  have been computed in
\cite{HT1}, where it was
found that the general coordinate 
and gauge  anomalies are canceled if the general coordinate and gauge 
transformations also assign a transformation of the antisymmetric
two-tensor, $B_{BA}=B_{BAx^{11}}$ 
obtained by dimensional reduction of the the 3-form $B_{BAC}$
\be
\delta B_{mn}=\oh \alpha'\{\epsilon^{AB}\partial_{[n}A_{m]}^{AB}-\oh\Theta^{ab}
 \partial_{[n}\om^{(-)ab}_{m]}\}
\ee
and modify the Bianchi identity for the torsion $H=dB$
\be
\partial_{[m}H_{npq]}=\frac{3}{16}\alpha'\{
F^{AB}_{[mn}F^{AB}_{pq]}-\oh R^{(-)ab}_{[mn}R^{(-)ab}_{pq]}\}\label{kkl}
\ee
where $\epsilon$ and $\Theta$ are the generators of the gauge and Lorentz
transformations \footnote{An anomaly in the general coordinate invariance is
equivalent to  anomaly in the 
Lorentz invariance \cite{HT1}.}.

So the world sheet anomaly cancellation leads to the 
same modified Bianchi identity for $H$ and  the same transformation 
for the two-tensor as those obtained by requiring spacetime supersymmetry and
the absence of anomalies  of 11 dimensional supergravity coupled to 
10 dimensional super Yang-Mills \cite{HW1}.

We should stress however, that the modification of the Bianchi identity
(\ref{kkl})  follows
from the preservation of the $\kappa$ symmetry, or alternatively of (0,1)
worldsheet supersymmetry, in close analogy with \cite{HW1}.  
As pointed out in \cite{Sen2}, the addition
of local counterterms to the effective action in order to 
restore gauge and Lorentz invariance breaks (0,1) supersymmetry
and therefore breaks $\kappa$ symmetry.  Fortunately, this term can be
supersymmetrized.  Alternatively \cite{HT1}, one can work in (0,1) superspace
to show that while preserving $\kappa$ symmetry one is able to restore
gauge and Lorentz invariance at the quantum level by using (\ref{kkl}).

The double dimensional reduction of the twisted supermembrane (\ref{S})
then follows the standard procedure of \cite{D}.  We 
use the reparametrization
invariance to set 
\ba
x^{11}&=&\rho.
\ea
We then require that 
\be
\partial_{\rho} Z^N=0\;\;N=1,...,10
\ee
and that 
\be
\partial_{x^{11}}\hat{G}_{MN}=\partial_{x^{11}}\hat{B}_{MNP}=0.
\ee
We can now express the eleven
dimensional variables in term of the ten dimensional ones  \cite{D}
\be
\hat{E}^A_M=\pmatrix{
{E}^a_M & {E}^{\alpha}_M+A_M \chi^{\alpha}& \Phi A_M \cr
0 &  \chi^{\alpha}& \Phi \cr
},
\ee
and
\be
\hat{B}_{MNP}=(B_{MNP},B_{MNx^{11}}=B_{MN}).
\ee
The superzenbein $E^A$ are right moving supersymmetric and have no left moving
fermions.

When $I$ is very small, the worldvolume is pure boundary.
The effective action we arrive to  in this limit is given 
by the boundary terms in action (\ref{S}), 
\ba
S_h&=&\int d^2\sigma
\{\oh (g_{mn}\eta^{ij}+b_{mn}\epsilon^{ij})\partial_ix^m\partial_jx^n\non\\
&&+\oh i \lambda^a_+[\partial_- \lambda^a_+
+\om_m^{(+)ab}\partial_-x^m\lambda^b_+]
\non\\
&&+\oh i \psi^A_-[\partial+ \psi^A_++A_{n}^{AB}\partial_+x^n\psi^B_-]
\non\\
&&+
\frac{1}{4}F_{abAB}\lambda^a_+\lambda^b_+\psi^A_-\psi^B_-\}
,\label{Shet}
\ea
and the gauge group indices $A$ and
$B$ now run over the $E_8\times E_8$ representation,
rather than the  $E_8$ representation.  The action (\ref{Shet})
is the usual heterotic string action  with $E_8\times E_8$ gauge group
and arbitrary background.  As expected, the theory also has a conformal
invariance which is also preserved at a quantum level.
Thus, after double dimensional  reduction of the non anomalous
twisted supermembrane action on $M^{10}\times I$, 
we obtain the $E_8\times E_8$ heterotic string.
This coincides with the expected low energy limit of M-theory \cite{HW}: 
the 11 dimensional 
supergravity action on $M^{10}\times I$
should yield upon dimensional reduction
the low energy theory for the heterotic string on $M^{10}$.

\section{ Heterotic-Type I duality}

We now consider the action $S$ on $M^9 \times S^1\times I$:
$E^A_i=\partial_i \hat{Z}^N \hat{E}^A_N$ and $\hat{E}^A_M$ is a superelfbein.
 $Z^N=(x^n,\theta^{\nu}),\ N=1,...,9$ are supercoordinates on $M^{9}$, $Z^{10}$ 
is a coordinate on $I$, and $Z^{11}$ 
is a supercoordinate
on $S^1$.

The double dimensional reduction of the supermembrane
 then follows the same 
standard procedure of \cite{D}.  We use the reparametrization
invariance to set 
\ba
x^{10}&=&\sigma \label{bc}\non\\
x^{11}&=&\rho
\ea
and require that 
\be
\partial_{\rho} Z^N=0\;\;N=1,...,10
\ee
and that 
\be
\partial_{x^{11}}\hat{G}_{MN}=\partial_{x^{11}}\hat{B}_{MNP}=0.
\ee
We can now express the eleven
dimensional variables in term of the ten dimensional ones  \cite{D}
\be
\hat{E}^A_M=\pmatrix{
E^a_M & E^{\alpha}_M+A_M \chi^{\alpha}& \Phi A_M \cr
0 &  \chi^{\alpha}& \Phi \cr
},
\ee
and
\be
\hat{B}_{MNP}=(B_{MNP},B_{MNx^{11}}=B_{MN}).
\ee
Before proceeding, we turn on Wilson lines on the boundaries of the 
supermembrane to break each $E_8$ to $SO(16)$.  
The motivation for such breaking of  the gauge group
is based on the fact that type IA string
will be the theory we will obtain
after double dimensional reduction of the supermembrane.  The type IA 
string will be anomaly free only for gauge group $SO(16)\times SO(16)$ 
\cite{H1}.
Therefore we
consider at each end of the membrane, gauge fields $A^{BC}$ such that
\ba
A^{BC}_{11}&=&\delta_{BC}, \;\;C=9,...,16,\non\\
A^{BC}_{m}&=&0\;\; \hbox{otherwise}
\ea
This means that the left moving fermions with $C=1,...,8$ are all massive.
Therefore, upon considering the limit in which 
$S^1$ is very small, the fermions with  $C=9,...,16$ do not contribute to
the effective action.

Thus, the effective action
 we arrive to when $S^1$ is very small is
\ba
S&=&\int d^2\zeta (\Phi \sqrt{-det({E}^a_i{E}^b_j\eta_{ab})}-\frac{1}{2}
\epsilon^{ij}{E}^A_i{E}^B_jB_{BA})\non\\
&&
+\int d\tau \sum_{A=1}^{8}\psi_0^A(\tau, +a)\partial_+
\psi_0^A(\tau, +a)\non\\
&&
+\int d\tau \sum_{B=1}^{8}\tilde{\psi}_0^B(\tau, -a)\partial_+
\tilde{\psi}_0^B(\tau, -a)
.\label{s1}
\ea
The first term in (\ref{s1}) is the Green-Schwarz action of the type I
string with Dirichlet boundary conditions for 
\be
x^{10}\in I.  \label{bc1}
\ee
The second and third terms  in (\ref{s1})  are similar to terms 
first proposed by Marcus and Sagnotti 
in \cite{MS} to construct open string with Chan Paton
factors \footnote{ I thank P. Ho\v{r}ava for pointing out this reference to
me}.  They
will provide only zero
modes, $\psi_0^A$ and $\psi_0^B$, with anticommutation relationship
\ba
\{\psi_0^A,\psi_0^B\}&=&2\delta^{AB},\non\\
\{\tilde{\psi}_0^A,\tilde{\psi}_0^B\}&=&2\delta^{AB}.
\ea
 This Clifford algebra, has a unique irreducible representation in terms of 
SO(8) Dirac spinors, one at each end of the open string.
The open  string now carries 16 dimensional indices
at each end or equivalently,
it carries an SO(16) representation at each
end.
Therefore, action (\ref{s1}) is just the action of the type IA string with
$SO(16)\times SO(16)$ gauge group discussed in \cite{H1,HW}.   
This is what is expected from 11 dimensional supergravity on 
$S^1\times I \times {M}^9$ in the limit in which the $S^1$ 
is very small \cite{HW}.
What we learn from the supermembrane analysis of 11 dimensional supergravity
is that the dimensional reduction of the
generators of the current algebra $E_8\times E_8$ of the
heterotic string, will yield the Chan Paton factors required
by the type IA string.

As explained in \cite{H1,H2}, it is possible to obtain the type I 
string with $SO(16)\times SO(16)$ from the type IA string.  
Type IA has Dirichlet boundary 
conditions (\ref{bc1}).  As shown in \cite{P}, under a T-duality transformation,
the Dirichlet boundary conditions are replaced by Neumann boundary conditions.
Thus, the type IA string is T-dual to the type I string. The type I string
thus obtained still has an $SO(16)\times SO(16)$ gauge group.  This means
that when $I$ vanishes, which is equivalent to a T-duality 
transformation of the type IA string, we recover the type I string with $
SO(16)\times SO(16)$ gauge group.

Alternatively, we could have considered the dimensional reduction of the 
$x_{10}$ coordinate instead of the $x^{11}$ coordinate of the supermembrane
action.  This would have lead to the heterotic string action with $
E_8\times E_8$ gauge group with $x^{11}\in S^1$.  We can break, once more,
using Wilson lines, the gauge group $E_8\times E_8$ to $SO(16)\times SO(16)$.
In the limit in which the
$S^1$ vanishes, which is equivalent to a T-duality transformation,
new massless fields appear and the whole SO(32) gauge symmetry of the
heterotic string is recovered
\cite{HW1,N,G}.
Thus, we recover the 
same result obtained in \cite{HW} by using the supermembrane.
This provides more evidence for the conjecture that the supermembrane
has as its low energy limit 11 dimensional supergravity.  Moreover,
the type of string we obtain is given by the choice of coordinates we decide
to dimensionally reduce.  This means that the underlying symmetry which is
responsible for string-string duality is the reparametrization invariance
of action (\ref{S}).  Using this symmetry, we can make the membrane look
like a thin cylinder, obtaining the open string, or we can make the membrane
look like a short cylinder, obtaining the heterotic string.

\section{Discussion}

This construction is 
an example where
the fixed points of a worldvolume manifold provide the  vector
multiplets which appear in the spectrum of the low energy effective action 
which is believed to be 11 dimensional supergravity on an orbifold.
This also provides evidence
for the
conjecture  made
in \cite{HW} that  the correct number of
vector multiplets appear in the spectrum
needed to cancel the gravitational anomaly of 11 
dimensional supergravity.  We summarize the supermembrane mechanism 
for generating  low energy effective actions with gauge group 
$SO(16)\times SO(16)$.
Cylindrical membranes, instead of toroidal membranes,  which
have boundaries require an $E_8$  current algebra at the boundaries
to cancel the 
3 dimensional worldvolume anomaly present at its boundaries.  This current
algebras supply the $E_8\times E_8$ current algebra of the heterotic string
when the cylinder collapses to a circle.  When the cylinder collapses to an
orbifolded circle, the current algebra present at the boundaries 
are the Chan Paton factors of the open string.  In order not to have
an anomalous type IA
theory, the $E_8$ current algebra must break down to $SO(16)$ at each
end \cite{HW}.

Perhaps the most surprising aspect of twisted supermembranes, is the
fact that the spacetime 
anomaly cancellation and supersymmetry requirements for
11 dimensional supergravity coupled to 10 dimensional $E_8$ super Yang-Mills
are the same requirements needed to cancel the gravitational anomaly of
 the twisted supermembrane.  This is more evidence supporting the 
idea that M-theory is a theory of closed and twisted membranes.

The perspective of M-theory being a theory of membranes, along with the
fact that twisted membranes lead to heterotic strings and close membranes
lead to type IIA string, suggests that membrane/string duality 
in seven dimensions \cite{T1}, or alternatively, string/string duality in
six dimensions, is actually  a duality between twisted and close membranes.

Another speculation is that the D-branes of M-theory 
can be constructed from the twisted membranes.
This follows from the fact that dimensional reduction of the twisted membrane
leads to an open string with Chan Paton factors at its ends, which are
present due to the algebra of the twisted 
membranes.  With this current algebra,
one can expect to construct low energy effective actions for the twisted
membranes of M-theory which have gauge fields, an essential ingredient
in the theory of D-branes.

\vspace{1cm}

{\bf Acknowledgements}

I am grateful to  P. Ho\v{r}ava for very useful discussions, 
and for
pointing out several aspects necessary for the completion of this article.
I should also like to thank A. Larsen,  J.D. Lewis, and B. Campbell.

\pagebreak


\begin{thebibliography}{10}


\bibitem{D} M.J.Duff, P.S.Howe, T.Inami, and K.S.Stelle, {\em Superstring 
in D=10 from supermembranes in D=11}, Phys. Lett. {\bf B191} (1987) 70.

\bibitem{T} P.K. Townsend, {\em The Eleven-Dimensional Supermembrane Revisited},
Phys. Lett. {\bf B350} (1995) 184.

\bibitem{T1} P.K.Townsend, {\em String-Membrane Duality in Seven Dimensions},
Phys. Lett. {\bf B354} (1995) 247.

\bibitem{HT} C.M.Hull and P.K.Townsend, {\em Enhanced Gauge Symmetries in
Superstring Theories}, Nucl. Phys. {\bf B451} (1995) 525.


\bibitem{HT1} C.M.Hull and P.K.Townsend, {\em World-Sheet Supersymmetry
and Anomaly Cancellation in the Heterotic String}, Phys. Lett. {\bf B178}
(1986) 187.

\bibitem{W1} E. Witten, {\em String Theory Dynamics in Various Dimensions},
Nucl. Phys. {\bf B443} (1995) 85.


\bibitem{HW} P. Ho\v{r}ava and E. Witten,
 {\em Heterotic and Type I String Dynamics
From Eleven Dimensions}, hepth 9510209.

\bibitem{HW1} P. Ho\v{r}ava and E. Witten, {\em Eleven-Dimensional Supergravity
on A Manifold with Boundary}, hepth 9603142.

\bibitem{H1} P. Ho\v{r}ava, {\em String on Worldsheet Orbifolds}, Nucl.Phys.
{\bf B327} (1989) 461.

\bibitem{P} J.Dai, R.G.Leigh, and J. Polchinski, {\em New Connections
Between String Theories}, Mod. Phys. Lett. {\bf A} (1989) 2073.

\bibitem{H2} P. Ho\v{r}ava, {\em Background Duality of Open String Models}, 
Phys. Lett. {\bf B231} (1989) 351.


\bibitem{Sc} J.H. Schwarz, {\em The Power of M theory}, Phys. Lett. {\bf B367} 
(1996) 97; \\
{\em An SL(2,Z) Multiplet of Type IIB Superstring}, Phys. Lett. {\bf B360}
(1995) 13.

\bibitem{GSW} M.B. Green, J.H. Schwarz and E. Witten, {\em Superstring
Theory:1}, Cambridge University Press, (1987).

\bibitem{ASY} O. Aharony, J. Sonnenschein, and S. Yankielowicz,
{\em Interactions of Strings and D-Branes from M-theory}, hepth 9603009.

\bibitem{TBS} E. Bergshoeff, E Sezgin, and P.K. Townsend, {\em Supermembranes
and Eleven-Dimensional Supergravity}, Phys. Lett. {\bf B198} (1987) 75;\\
{\em Properties of 
the Eleven-Dimensional Supermembrane Theory,}  Ann. Phys. {\bf 185} (1988) 330.

\bibitem{Sen} A. Sen, {\em String String Duality Conjecture In Six Dimensions
and Charged Solitonic Strings }, Nucl. Phys. {\bf B450} (1995) 103.

\bibitem{Sen2} A. Sen, {\em Local Gauge and Lorentz Invariance of the Heterotic
String Theory}, Phys. Lett. {\bf B166} (1986) 300.



\bibitem{DW} B. De Wit, M. Luscher and H. Nicolai, {\em 
The Supermembrane is Unstable},
Nucl.Phys. {\bf B320}
(1989)
135.


\bibitem{AL} F. Aldabe and A. Larsen, {\em 
Supermembranes and Superstrings with Extrinsic
Curvature}, hepth 9602112.

\bibitem{MS} N. Marcus, and A. Sagnotti, {\em Group Theory From 
Quarks at the Ends of Strings}, Phys. Lett. {\bf B188} (1987) 58.


\bibitem{N} K.S. Narain, {\em New Heterotic String Theory In Uncompactified
Dimensions <10}, Phys. Lett. {\bf B169} (1986) 41; K.S. Narain,  M.H. Sarmadi,
and E. Witten, {\em A Note on Toroidal Compactifications of Heterotic String
Theory}, Nucl. Phys. {\bf B279} (1987) 369.

\bibitem{G} P. Ginsparg, {\em On Toroidal Compactification of Heterotic 
Superstrings}, Phys. Rev {\bf D35} (1987) 648.

\end{thebibliography}
\end{document}